\documentstyle[12pt]{article}
\newcommand{\draft}{%           Normal spacing
        \renewcommand{\baselinestretch}{1.0}%
        \small\normalsize%
}

\draft
\begin{document}
\title{\bf The Lower Limit for Masses of Progenitors of Supernova 
Remnants and Radio Pulsars} 
\author{Sevin\c{c} O. Tagieva$\sp1$   
\thanks{e-mail:physic@lan.ab.az},
Oktay H. Guseinov$\sp{2,3}$
\thanks{e-mail:huseyin@gursey.gov.tr},
A\c{s}k\i n Ankay$\sp2$
\thanks{email:askin@gursey.gov.tr}, \\ \\
{$\sp1$Academy of Science, Physics Institute, Baku 370143,} \\
{Azerbaijan Republic} \\
{$\sp2$Bo\u{g}azi\c{c}i Uni. - T\"{U}B\.{I}TAK Feza G\"{u}rsey Institute,}
\\
{81220 \c{C}engelk\"{o}y, \.{I}stanbul, Turkey} \\
{$\sp3$Akdeniz University, Department of Physics,} \\
{Antalya, Turkey}}
%\\ \\ \\ \\ \\ 
%{3 copies of: ?? pages of text + 4 pages of Tables.} \\ \\
%{Please send proofs to:} \\
%{A\c{s}k\i n Ankay} \\
%{Feza G\"{u}rsey Enstit\"{u}s\"{u},} \\
%{81220 \c{C}engelk\"{o}y,} \\
%{\.{I}stanbul,} \\
%{Turkey} \\
%{(e-mail: askin@gursey.gov.tr)}}

\date{} 
\maketitle 
%\final 

\begin{abstract} 
\noindent 
We examined correlations between young radio pulsars (PSRs), Supernova
remnants (SNRs) which have different surface brightness values and young
star formation regions (SFRs). Angular correlation of PSRs with SFRs is
reliable up to 4 kpc and considerably strong up to 3 kpc from the Sun. 
On average this correlation is stronger for Galactic anticenter directions 
compared to Galactic central directions. Angular correlation of SNRs with 
SFRs is weaker and depends on the surface brightness of the SNR. 
Spatial correlation of PSRs with SFRs is also stronger than spatial 
correlation of SNRs with SFRs. Dim SNRs show weak spatial correlation with 
SFRs. These investigations and analysis of various data show that the 
lower limit for masses of progenitors of PSRs is about 9 M$_{\odot}$ and 
of SNRs (or supernovae) is about 7 M$_{\odot}$.
\end{abstract}

%Key Words: radio pulsar, supernova, supernova remnant, white dwarf, 
%progenitor mass

\clearpage
\parindent=0.2in

\section{Introduction}
Although the problem of masses of the progenitors (on the main sequence) 
of radio pulsars (PSRs) and Supernova remnants (SNRs) has been 
discussed for many years, there are still open questions. 
It is not well known under which circumstances a neutron star (NS) is  
born during a Supernova (SN) explosion. Actually, PSRs and in general NSs 
do exist in SNRs which are formed as a result of SN explosions with 
energies 10$^{49}$-10$^{51}$ erg, in some cases even with energies 
$<$10$^{49}$ erg (e.g. Crab $^1$) and with energies $>$10$^{51}$ erg 
(e.g. Cas A $^{2,3}$). How can we determine the lower limit for masses of 
NS and PSR progenitors and masses of the stars for which the evolution 
ends with SN explosion (excluding the accretion induced collapse)?

It was claimed that PSRs are formed once every 150 years
and the lower limit for masses of the stars which form PSRs at the end
of their evolution is about 5 M$_{\odot}$ $^4$. On the other hand, by 
examining the historical SNRs it was claimed that a SN occurs every 6 
years and the lower 
limit for the mass of the progenitors is also about 5 M$_{\odot}$ $^5$.
Does the formation of PSRs predominantly depend on some other parameters
$^6$ as the lower limit for the progenitor mass is 5
M$_{\odot}$ in both cases?

In principle, the lower limit for masses of the progenitors of the 
stars, which experience SN explosion and possibly form PSRs and in general 
NSs, may be equal to the higher limit for masses of the progenitors of 
single white dwarfs (WDs). According to a theoretical calculation  
$^7$, the higher limit for masses of the progenitors of WDs is about
7-8 M$_{\odot}$. This higher limit was also given as 6-11 M$_{\odot}$ 
$^8$. Today, there is no observational data showing that mass 
of the progenitor of a WD can be $>$6-7 M$_{\odot}$; this result comes 
from the investigations of WDs in young open clusters (OCs). WD 0349+247,
which has a mass of $\sim$1 M$_{\odot}$ in the OC Pleiades, has a
progenitor with a mass of $\sim$6 M$_{\odot}$ $^{9,10}$.

Now, lets briefly look at the perspective to find the higher mass limit
for progenitors of WDs in OCs. For a star with 7 M$_{\odot}$ the lifetime 
is $\sim$4$\times$10$^7$ yrs $^{11}$, but the uncertainties
in cooling times (i.e. ages) of WDs with high temperatures ($\sim$30000 K) 
are greater than the lifetime of the progenitor (only such WDs are 
observable at large distances). If we take into account the distance 
of the young OC
NGC 2168 (870 pc), we can see how difficult it is to find the upper limit
for the mass from the observations of WDs in this and other distant OCs.
One must add up the errors in determinations of temperatures and masses   
of WDs with the errors in WD cooling models. So, it is more reliable, but
not easier, to search for planetary nebulae and their stellar remnants in
young OCs with turn-off masses $\sim$6-8 M$_{\odot}$ $^{12}$. By this way 
we get rid of the difficulties mentioned above. We can assume that the 
mass of the progenitor is equal to the turn-off mass for the stars in the 
OC, since the lifetime of planetary nebulae is very short 
($\sim$10$^4$-10$^5$ yrs).

Assuming the lower limit for the mass of progenitors of SN
to be 5 or 8 M$_{\odot}$ leads to a difference of a factor of 3 in the 
formation rate of SNRs, if we use the initial mass function (IMF) of 
Blaha \& Humphreys $^{13}$. In the Solar neighbourhood the uncertainty in 
the IMF up to 
$\sim$15-16 M$_{\odot}$ is not large. Even for different galaxies and star 
formation regions (SFRs) we may use a simple IMF, similar to the one 
given by Salpeter $^{14}$, with a value of power about 2.3--3 $^{15}$. 

The problem mentioned above is an actual one and it is essential to try to
find reliable ways to understand the differences in the birth rates of  
SNR, PSR and NS. But there are some difficulties: there are significant 
uncertainties in the data because of taking large volumes and long time 
intervals into account in order to find a statistical result, 
but even if we try to increase the number of the statistical data by this 
way, we can not have enough number of them to be used in order to obtain 
reliable results in statistical investigations. So, one must use all the 
possible independent ways to solve the problem. In fact, neither of these 
ways leads to a reliable result by itself because of small number of data 
even if it leads to a result which is actually true. On the other hand, we 
can compare the results of these independent ways and try to find a 
reliable result. So, we can determine the lower limit for the mass of 
progenitors of PSRs, NSs and SNRs in this way; we will use independent 
data such as the upper limit for masses of the progenitors of single WDs, 
formation rates of SNe based on IMF and observationally found SN rate for 
Sb type galaxies together with the relations between the locations of 
young PSRs, SNRs, young SFRs and OB associations in the Galaxy. 
\section{Connections of SFRs to PSRs and SNRs: the general view}
Average space velocity of PSRs is about 250-300 km/s $^{16,17}$
so that, PSRs with large ages can go very far away from their birth 
sites (i.e. from OB associations and partly 
from SFRs). Since we want to examine the relations of PSRs and SNRs with
their birth sites, we will take into account only the PSRs which  
have characteristic ages $\tau$ $\le$ 10$^6$ yrs. On average, PSRs 
with such ages can not go more than 200 pc far away from the Galactic 
plane. There are 259 PSRs with $\tau$ $\le$ 10$^6$ yrs, $\mid$b$\mid$ $<$ 
5$^o$, and F$_{1400}$ $\ge$ 0.2 mJy values $^{18}$. Sixty of these PSRs 
are located within 4 kpc around the Sun $^{19}$. Below, we will use this 
sample of PSRs.

About 75\% of O-type stars are members of OB-associations, whereas, only 
about 58\% of the stars earlier than B3-type (i.e. the stars with masses 
$\ge$9 M$_{\odot}$) are members of OB-associations $^{20}$. On the other 
hand, in the region with a radius of 110 pc 
in the Solar neighbourhood, there are 6 stars having masses $>$9 
M$_{\odot}$ $^{21}$. It is known that massive stars which are 
not members of OB-associations and OCs are located in SFRs. The Sun is 
also located in a SFR and there is a surrounding environment with young 
OCs (Perseus, Pleiades, $o$ Vela and Carina) which contain 6-7 M$_{\odot}$ 
stars in its near neighbourhood. A bit farther away, at a distance of
180 pc, there is OB-association SCO OB2. Because of these facts, we can  
assume that $\ge$50\% of the stars with masses 7 M$_{\odot}$ $<$ M $<$ 9
M$_{\odot}$ and most of the stars which have masses $\ge$9 M$_{\odot}$  
are located in SFRs. The SFRs are actually much wider than OB-associations 
$^{22}$.

In spite of the considerable influence of HII regions on lifetimes of 
SNRs, the distributions of PSRs and SNRs must be correlated with the 
distributions of SFRs. In addition to this, if the lower limits for the 
mass of the progenitors of these objects are considerably different from 
each other, then there will be a weak correlation between PSRs and SNRs. 

Distance versus Galactic longitude diagram of the PSRs with 
characteristic ages $\tau$$\le$10$^6$ yr and $\mid$b$\mid$ $<$ 5$^o$ 
is displayed in Figure 1. By comparing Figure 1 with the Galactic arms 
and the distribution of HII regions $^{23}$ we see that in general the 
distribution of such young PSRs is in accordance with the Galactic arm 
structure. Despite the fact that there is no arm nor subparts of an arm in 
the Galactic longitude interval l = 270$^o$ $\pm$ 10$^o$ $^{23}$, PSRs are 
located densely in these directions, particularly at 5-6 kpc. In 
actuality, Vela I-IV associations are located in these directions that 
there is a large SFR in this direction. Among these OB-associations the 
nearest one, Vela OB2, has a distance of $\sim$0.5 kpc and the most 
distant one, Vela OB1, is located at $\sim$2 kpc $^{24,25}$. In these 
directions, there are many young OCs at different distances $^{26,27}$
and there are many red supergiants at
5-6 kpc $^{28}$. As seen from Figure 1, the number of PSRs is 
also large in the intervals d=2-3 kpc and d=6-10 kpc in the direction   
l=300$^o$. In this direction, first the Carina arm and then both the  
Carina arm and a subpart of it are cut at about the same distance values. 
There is a large region which does not contain any PSR in this direction 
at distances $\sim$3-4 kpc (see Figure 1) and this shows that there is 
actually no SFR between the arms Carina and Sagittarius. As seen in 
Figure 1, there are many young PSRs roughly in the interval 275$^o$ $<$ l 
$<$ 325$^o$ at about 10-15 kpc. This shows that another arm or a large 
subpart of Carina arm is passing through this region. The number of PSRs 
being large in the interval l = 25$^o$ $\pm$ 5$^o$ (Figure 1) fits  
well to the distributions of the giant HII regions $^{23}$.

In Figure 2, distance versus Galactic longitude diagram of 208 SNRs which 
have $\mid$b$\mid$$<$5$^o$ is represented (the distance values were taken 
from Guseinov et al. $^{29}$). When we compare Figure 2 with the arm 
structures and the distribution of the HII regions, we see that the SNRs 
and O-type stars do not show clustering likewise. 
\section{Angular correlations between the PSRs, SNRs and the SFRs with 
many massive stars} 
Let's now examine the relations between the places of young PSRs, SNRs 
and SFRs around the Sun. As known, depending on the distance (beyond 
$\sim$3 kpc) the selection effect in finding O-type stars and their 
associations in the Galactic 
central directions becomes very significant. But considering the fact
that the numbers of the SNRs and the PSRs are small, first
we will take into account only 29 bright SNRs with $\Sigma$$>$10$^{-21}$
Wm$^{-2}$Hz$^{-1}$sr$^{-1}$ and d$\le$4 kpc, and 60 PSRs with
$\tau$$\le$10$^6$ yr, $\mid$b$\mid$$<$5$^o$, F$_{1400}$$\ge$0.2 mJy and  
d$\le$4 kpc. Today, the SNRs with such surface brightness values and 
the PSRs with such fluxes are directly observable also in the Galactic 
central directions $^{30}$. The number of PSRs in this sample
is small in spite of the existence of $\sim$900 PSRs with measured
F$_{1400}$ values. The flux values of nearby PSRs are known, but,       
independent of the flux value, there are only 60 PSRs satisfying  
the criteria d$\le$4 kpc, $\mid$b$\mid$$<$5$^o$ and $\tau$$\le$10$^6$ yr.
Below, we will examine projections of the PSRs and the SNRs in our   
sample on the SFRs.

Among the PSRs in our sample some of them have projections on the SFRs 
$^{24,25,31,32,33}$ which include many massive stars (i.e. O-type stars
and supergiants). Numbers of such PSRs are given in Table 1. Forty-eight 
PSRs are located in the Galactic central directions. Six out of 9 SFRs 
given in Table 1 are also in the Galactic central directions. 
We have chosen the Galactic latitude $\mid$b$\mid$ intervals for each SFR
considering the minimum and the maximum latitude values of the
OB-associations (and the OCs) for each region. Only 17 of the PSRs
in our sample (6 of which have ages $\le$10$^5$ yr that there is no
difference in the ages of the PSRs which are projected on the SFRs and 
which are not) have projections on 6 of the SFRs in the Galactic central 
directions given in Table 1. As seen from Table 1, 12 out of 17 PSRs which 
are in the interval --9$^o$ $<$ l $<$ 14$^o$ are projected on the SFRs. 
The number of SNRs which have projections on the SFRs in the central 
directions is 3 (Table 1). On the other hand, the number of bright SNRs in
the Galactic central directions is 20 in our sample. From these data, the
portions of the PSRs and the SNRs which are projected on the SFRs are
0.35 and 0.15, respectively, whereas the ratio of the total area of these 
SFRs to the area of the sky with l=0$^o$$\pm$90$^o$ and 
$\mid$b$\mid$$\le$5$^o$ is 0.13. So, most of the PSRs are born in SFRs, 
whereas, the SNRs seem to have weaker connections with SFRs. 
But the value of the angular correlation must be more because of 
a selection effect which we can not take into account quantitatively; it 
is well known that the number density of even very young objects increases 
in the central direction. The maxima of the number density distributions 
are within $\sim$3-4 kpc of the Galactic center. Therefore, there exist 
many SFRs (which we did not take into account) in the region up to 4 kpc 
from the Sun in the Galactic central directions. A similar effect, but in 
smaller degree, exists also in the directions 60$^o$ $<$ l $<$ 300$^o$.

There are totally 29 SNRs which are located within 4 kpc around the Sun in
the Galactic longitude interval 60$^o$ $<$ l $<$ 300$^o$ with surface
brightness values $\Sigma$$>$3$\times$10$^{-22}$ 
Wm$^{-2}$Hz$^{-1}$sr$^{-1}$ (here we can neglect the influence of the 
background radiation $^{30}$). Numbers of the SNRs with 
$\Sigma$$>$3$\times$10$^{-22}$ Wm$^{-2}$Hz$^{-1}$sr$^{-1}$ projected on 
the SFRs in this longitude interval are also shown in Table 1.

Ten of the PSRs are projected
on the 5 SFRs (see Table 1) and the total number of PSRs included in
our sample in these directions is 19. Eight out of the 29 SNRs with
$\Sigma$$>$3$\times$10$^{-22}$ Wm$^{-2}$Hz$^{-1}$sr$^{-1}$ in the 
directions 60$^o$ $<$ l $<$ 300$^o$ have
projections on these 5 SFRs. Therefore, the ratios of the numbers of the
PSRs and the SNRs projected on these SFRs to the numbers of the PSRs and
the SNRs in the interval 60$^o$ $<$ l $<$ 300$^o$ respectively are 0.53   
and 0.30. The ratio of the total area of the SFRs in these 
directions to the area of the sky with 60$^o$ $<$ l $<$ 300$^o$ 
and $\mid$b$\mid$$\le$5$^o$ is 0.21. Therefore, the number of PSRs 
projected on the SFRs in these directions is considerably larger than the 
expected number of chance projections, but the number of the SNRs 
projected on the SFRs is a bit larger than the expected number of chance 
projections. Since Vela SNR and PSR has b=--2.74$^o$, they do not seem to 
be projected on SFR. A similar situation also arises for Cas A. But these 
objects have relations with SFRs, therefore, we have included them in the 
statistics.   

It is necessary to note that the HII regions in each large 
SFR have no effect on the searches of the PSRs and the SNRs in our sample 
$^{30}$, but the effect of HII regions on the SNRs is not
only related to the background. The SNRs which are born at the end of 
evolution of massive stars expand in HII regions and they have short 
lifetimes $^{34,35}$. On the other hand, the 
southern hemisphere was not searched so well compared to other regions. 
Although, there are 13 PSRs projected on the last 4 SFRs shown in 
Table 1, there is no SNR projected on these SFRs.

As seen from Table 1, SFRs are formed by OB-associations and OCs (e.g. SGR 
OB1 forms the first SFR in Table 1, whereas, the second SFR contain 3 
OB associations, namely, CYG OB3, CYG OB1 and CYG OB9).
The sum of longitude intervals of OB associations (and OCs) is
close to the sum of the longitude intervals of SFRs in which they are
included (see Table 1). We have not taken into account the overlapping 
parts, so that, the difference between the sum of angular sizes of OB 
associations and of the SFRs arises from the latitude intervals. 
Therefore, the sum of areas of SFRs may be up to 2-3 times more than the 
sum of areas of OB associations (the Vela being a special case). On the 
other hand, since progenitor stars and especially
PSRs with d$<$4 kpc may change their positions more than 1$^o$, it is
difficult to expect that angular correlation between SNRs, PSRs and some  
of the OB associations can give an additional result. The PSR which has
projection on the second SFR (see Table 1) is also projected on the OB
association CYG OB9 which contain a considerably small number of massive
(O-type stars and in general the stars with M$>$7 M$_{\odot}$) stars 
compared to the other 2 OB associations in the SFR $^{33}$. Among the 5 
SNRs which have projections on the third SFR, 2 of them 
are projected on CAS OB5 and one of them is projected on CAS OB4. Also 
one of the PSRs is projected on CAS OB5. None of the SNRs nor PSRs has a
projection on PER OB1 in which the number of massive stars is several 
times more compared to each of the other OB associations in the SFR. 
It is necessary to take into consideration that, in this direction, in the
intervals 128$^o$ $<$ l $<$ 155$^o$ and $\mid$b$\mid$ $<$ 6$^o$, there  
are only 2 SNRs and 4 PSRs (independent of any parameters). 

Among the 4-5 SNRs in our sample which have projections on the OB
associations given in Table 1, SNR G116.9+0.2 (CTB1) has the lowest 
surface
brightness ($\Sigma$=1.17$\times$10$^{-21}$ Wm$^{-2}$Hz$^{-1}$sr$^{-1}$) 
and this SNR has a projection on CAS OB5 association. This SNR is
expanding in a supercavity $^{36}$ and since it is located at
3.5 kpc $^{29}$ it has no genetic connection with CAS     
OB5. SNR G78.2+2.1 may be expanding in a cavity $^{37}$.
SNR 89.0+4.7 (HB 21) has a connection with CYG OB7 $^{38}$
(not included as a SFR in our work). The last one is SNR G120.1+1.4 
(Tycho). The distances of the last 3 SNRs are comparable with the 
distances of the related OB associations, so that, they can be
physically connected to Cas OB4. All of these 3 SNRs have 
$\Sigma$$>$3$\times$10$^{-21}$ Wm$^{-2}$Hz$^{-1}$sr$^{-1}$, so that, the 
small sizes and ages are in agreement with the conclusions of 
Lozinskaya $^{34}$ and Tagieva $^{35}$. 

\section{Spatial correlation of PSRs and SNRs with SFRs around the Sun at
distances up to 3 kpc}
In Figure 3, spatial distributions of the OB associations $^{24}$ 
for the regions d$\le$3 kpc,
60$^o$ $<$ l $<$ 300$^o$ and d$\le$2.5 kpc, l=0$^o$$\pm$60$^o$ are 
displayed. All the PSRs with $\tau$$\le$10$^6$ yr and F$_{1400}$$>$0.2 mJy 
in the same regions are also represented in this figure. 
Since the number of the PSRs is small, the 
boundaries of the SFRs are determined roughly $^{22}$, and the 
uncertainties in the distance values are comparable with sizes of the 
SFRs, it is not easy to find an exact value for the portion of the PSRs 
genetically connected to the SFRs. In the figure, there are 20 PSRs in 
the interval 60$^o$ $<$ l $<$ 300$^o$. We can roughly assume that about 
11-13 of these PSRs (i.e. $\sim$55-65\% of them) are genetically 
connected to SFRs. The total number of PSRs shown in Figure 3 is 31; 16-18 
of them can be assumed to have genetic connections with 
the SFRs. Please note that there is Persei arm in the direction 
l$\sim$120$^o$ at about 3 kpc and there are many HII regions there. So, 
the PSRs in this region must have connections with SFRs. All of these show 
that about 55-60\% of the PSRs are connected to SFRs. On the other hand, 
projections of the SFRs on the plane of the Galaxy in this region occupy 
approximately 30\% of the area of the region under consideration (Figure 
3).

In Figure 4, spatial distributions of the same OB associations, the  
SNRs with 3$\times$10$^{-22}$ $<$ $\Sigma$ $\le$ 10$^{-21}$
Wm$^{-2}$Hz$^{-1}$sr$^{-1}$ and the SNRs with $\Sigma$ $>$ 10$^{-21}$ 
Wm$^{-2}$Hz$^{-1}$sr$^{-1}$ in the same regions at the same distances of 
the PSRs are displayed. It is seen that out of the 17
bright SNRs 9 of them are genetically connected to the SFRs. The   
historical SNRs Tycho, Cas A and SN1181 are located at distances a little
more than 3 kpc (Figure 4). We do not show the locations of OB
associations beyond 3 kpc, but it is known that all of these SNRs are
located in SFRs. If we exclude these 3 remnants, fraction of the SNRs
which have connections with SFRs will be 6 out of 14.
Out of the 9 dim SNRs only 3-4 of them may be connected to the SFRs. 
There are also 10 SNRs with $\Sigma$ $<$ 3$\times$10$^{-22}$
Wm$^{-2}$Hz$^{-1}$sr$^{-1}$ located in the considered
region (60$^o$ $<$ l $<$ 300$^o$). These 10 SNRs were all observed in the 
interval 60$^o$ $<$ l $<$ 240$^o$ because this interval was 
observed more precisely and with high sensitivity. Among these 10 SNRs,
3-4 of them may be connected to SFRs. As seen from Figure 4, about 
30-45\% of the SNRs in our sample can have genetic connections with 
SFRs. But we must also take into account short lifetime of the SNRs in 
HII regions. Therefore, the fraction of SNRs which have connections 
with SFRs must be about 45\%. Please note that the connections would be 
$\sim$30\% for the chance distribution of these objects.
\section{Discussion and Conclusions}
Observational data of PSRs $^{18}$ 
and SNRs $^{39}$ show that even in the Galactic central
directions (l=0$^o$$\pm$10$^o$, $\mid$b$\mid$$<$2$^o$) all the SNRs with
$\Sigma$$>$10$^{-21}$ Wm$^{-2}$Hz$^{-1}$sr$^{-1}$ and the PSRs with   
F$_{1400}$$>$0.2 mJy are observable. Also, the SNRs in the interval 60$^o$ 
$<$ l $<$ 300$^o$, for all values of b, can be easily observed if
$\Sigma$$>$3$\times$10$^{-22}$ Wm$^{-2}$Hz$^{-1}$sr$^{-1}$ $^{30}$.

We have taken this into account and in Section 3 we have examined the 
angular correlations between the PSRs with $\tau$$\le$10$^6$
yr, $\mid$b$\mid$$<$5$^o$, F$_{1400}$$\ge$0.2 mJy, the
SNRs which have $\Sigma$$>$10$^{-21}$ Wm$^{-2}$Hz$^{-1}$sr$^{-1}$, 
$\mid$b$\mid$$<$5$^o$ and $\Sigma$$>$3$\times$10$^{-22}$ 
Wm$^{-2}$Hz$^{-1}$sr$^{-1}$, $\mid$b$\mid$$<$5$^o$, and the 
SFRs rich in O-type stars, which are located up
to 4 kpc. The correlation of the PSRs in the Galactic central directions
(l=0$^o$$\pm$90$^o$) with the SFRs is strong. The probability of chance
projection of the PSRs in our sample on the SFRs in these
directions is 13\%, but in actuality the percentage of the ones which   
have projections is not smaller than 35\% if we take into account 
high space velocities of PSRs. In the Galactic
longitude interval 60$^o$ $<$ l $<$ 300$^o$ the probability of chance
projection is 21\%, but actually $\ge$53\% of the PSRs are projected on 
the SFRs.

The angular correlation of the SNRs in the Galactic central directions
with the SFRs is 15\%, whereas, the chance projection is 13\%, but  
number of the statistical data is very small. The probability 
of projection on the SFRs for the SNRs with $\Sigma$$>$3$\times$10$^{-22}$ 
Wm$^{-2}$Hz$^{-1}$sr$^{-1}$ and $\mid$b$\mid$$<$5$^o$ in the
interval 60$^o$ $<$ l $<$ 300$^o$ is 30\%, i.e. 1.4 times larger than
the distribution by chance. 
Although PSRs and SNRs are born due to SN explosions, there is no angular
correlation for the spatial distributions of young PSRs and the SNRs in
our sample, either. This must be the result of different precisions and
sensitivities of the PSR and SNR searches in different directions of the
Galaxy as well as short lifetimes of the SNRs which are born at the end of 
the evolution of O-type stars. For example in the sector 270$^o$ $<$ l 
$<$ 360$^o$ there is a large number of PSRs and a small number of 
SNRs, while in the region 60$^o$ $<$ l $<$ 90$^o$ number of SNRs is 
large (see Figures 3 and 4). Another reason can be that the average 
mass of progenitors of PSRs is a bit larger than the masses of 
progenitors of SNRs.

We have considered the same correlation also for the PSRs and the SNRs in 
our sample and the SFRs all of which are located up to 3 kpc. We have 
found the value of 45\%, instead of 35\%
found for the PSRs with d$\le$4 kpc in the central directions. For the   
interval 60$^o$ $<$ l $<$ 300$^o$, we have found the percentage of the 
PSRs which have projections on the SFRs to be 64\% instead of 53\% which 
was found for d$\le$4 kpc case. On the other hand, for the SNRs in the 
Galactic central directions, instead of the 
value of 15\% found for d$\le$4 kpc case, we have found the value of 18\%. 
For the SNRs in the interval 60$^o$ $<$ l $<$
300$^o$ we have found a value of 22\% instead of the value of 30\% found
for the d$\le$ 4 kpc case. So, we can assume that the actual projections 
of young PSRs on the SFRs is about 3 times larger than the chance 
projections. If we consider the short lifetime of SNRs in HII regions, 
then the actual projections of the SNRs on the SFRs will be about 1.5 
times more than the chance projections.

The increase of the correlation under the diminishing of the volume should 
be considered normal. The PSRs and SNRs in our sample can be observed 
easily up to 4 kpc, but it is difficult to observe O-type stars beyond 2-3 
kpc, especially the OB associations in the central directions. Therefore, 
the most distant PSRs and SNRs actually may have projections on the SFRs 
which were not observed. Existence of such an effect must show itself 
also for O-type stars, but weakly. The diminishing of the percentage of 
the SNRs, which have projections on the SFRs, at smaller distances show 
that the SNRs actually have weaker correlation with young SFRs compared 
to the PSRs. 

Is the angular correlation of the O-type stars given in 
Cruz-Gonzalez et al. $^{40}$ with the 
SFRs in the region 60$^o$ $<$ l $<$ 300$^o$ much
different compared to the correlation with the PSRs? There are 309 O-type
stars in the directions of these SFRs and the total number of O-type stars
in this region is 455, so that, the angular correlation is 68\%. This is 
possibly an underestimated value because we do not put a limit for the 
distances of O-type stars as we have put for the SFRs. But still this 
value is a bit larger than the correlation of the PSRs located less than 
3 kpc and significantly larger than the correlation of the PSRs with 
d$\le$4 kpc. This shows that PSRs have, in general, progenitors with 
masses $>$9 M$_{\odot}$ (i.e. earlier than B3V-type stars). 
%This must be true, because the correlation of the PSRs with the SFRs is 
%close to the correlation of O-type stars with the SFRs in this region. 
It is necessary to remember 
that about 75\% of O-type stars are members of OB associations, whereas, 
only about 58\% of the stars earlier than B3-type are members of OB 
associations $^{20}$.

We have tried to find relations between the spatial distributions of the 
SFRs on the basis of the OB associations classified by Melnik \& Efremov 
$^{24}$ and of the PSRs with characteristic ages $\tau$$\le$10$^6$ yr and 
F$_{1400}$$>$0.2 mJy, of the SNRs with $\Sigma$$>$10$^{-21}$ 
Wm$^{-2}$Hz$^{-1}$sr$^{-1}$, and of the SNRs with 3$\times$10$^{-22}$ $<$ 
$\Sigma$ $<$ 10$^{-21}$ Wm$^{-2}$Hz$^{-1}$sr$^{-1}$. We have compared the 
spatial distributions for the
interval 60$^o$ $<$ l $<$ 300$^o$ up to 3 kpc from the Sun, taking into
account that OB associations were identified better in this angular
interval. On the other hand, for the region l=$\pm$60$^o$ in the central
direction, we have considered the correlations of the objects located 
up to 2.5 kpc. We attempt to take into account the
uncertainties in the distance values of these objects and the high space
velocities of PSRs. Naturally, the errors in the determination of sizes of
the SFRs together with the numbers of the PSRs and SNRs being small make 
the uncertainties larger. This does not change the reliability of
whether there is a correlation or not, but has some effect on the 
quantitative value of the correlation. About 55-60\% of the young PSRs  
and about 45-55\% of the SNRs with
$\Sigma$$>$10$^{-21}$ Wm$^{-2}$Hz$^{-1}$sr$^{-1}$ are genetically
connected to the SFRs, whereas only about 30-40\% of the dimmer SNRs have
connections with the SFRs. As seen from Figures 3 and 4, sum of the areas 
of the SFRs is about 30\% of the total area chosen, so that, the 
connections between these objects must be $\sim$30\% under chance 
distributions.

Actually, there are some O-type stars, other than the ones in the SFRs
shown in Figure 3, included in the catalogue of Cruz-Gonzalez et 
al. $^{40}$, particularly around Crab PSR. Taking this and especially the
discussions in this section into account, we can say that most of the   
progenitors of PSRs and in general of NSs are more massive (like the
progenitors of most of the bright SNRs) compared to the progenitors of
dim SNRs. This is also supported by a comparison of the distance versus 
longitude distributions of young PSRs and SNRs (Figures 1 and 2) with the 
distribution of HII regions and the Galactic arm structures $^{23}$. 

Only SNR Tycho among the 6 historical SNRs (age $\le$ 1000 yr) is
probably a member of an OB association. But if all of the progenitors of
SNe had masses $>$ 9 M$_{\odot}$, then most of the SNRs and PSRs would be
born (but not necessarily located) in or near OB associations. Although,
SNRs Cas A and SN1181 are not members of OB associations, they are located
in SFRs. None of the other historical SNRs (Kepler, Crab and SN1006) was 
found to be in OB association nor in or close to SFR, even if they were 
searched in all wavelength bands.

According to Ankay et al. $^{30}$, on average, one SN explosion occurs 
every 65 yr and one PSR is formed every 300 yr in the Galaxy. 
On the other hand, the upper limit for masses of the progenitors of WDs 
is $\sim$7 M$_{\odot}$. This value is also the lower limit for masses of 
the progenitors of SNe, if the lower limit and the upper 
limit do not overlap. If we use this value of lower limit together with 
the IMF, we find the formation rate of SNe to be $\sim$1-2 in 100 years, 
which is in accordance with the observations $^{41}$. If the lower limit 
for masses of the progenitors of SNe were less than $\sim$7 M$_{\odot}$, 
then the formation rate of SNe would be higher and there would not be 
spatial correlation between the SNRs and the SFRs. Please note that the 
SFRs considered in this work include OB associations and very young OCs 
which contain O-type stars. Since lifetimes of the stars with masses $<$7 
M$_{\odot}$ are large, we do not expect spatial correlation between such 
stars and the SFRs used in this work. So, we can say that the 
lower limit mass for the progenitors of SNe and SNRs is $\sim$7 
M$_{\odot}$. From the spatial correlation of young PSRs with the SFRs we 
see that the lower limit for masses of the progenitors of PSRs must be 
$\sim$9 M$_{\odot}$. The difference between the lower limits for the 
masses of progenitors of SNe and PSRs ($\sim$7 M$_{\odot}$ and $\sim$9 
M$_{\odot}$ respectively) does not correspond to the ratio of
the birth rates of these objects which is about 4.5 $^{30}$. 
So, we can expect some additional conditions for the formation of PSRs, 
possibly an idea similar to the one claimed by Iben \& Tutukov $^6$. \\ 
{\it Acknowledgment:} We thank Efe Yazgan and M. \"{O}zg\"{u}r Ta\c{s}k\i 
n for their help.

%\clearpage
\draft
\begin{flushleft}
\begin{tabular}{cccccccc}
\multicolumn{8}{l}{\bf Table 1 \hspace{0.2cm} Clusters of OB associations
and young ($<$10$^8$ yr) OCs} \\
\multicolumn{8}{l}{\bf and the numbers of SNRs and PSRs in our sample} \\
\multicolumn{8}{l}{\bf which are projected on the chosen clusters} \\
\hline
Name & $\Delta$l($^o$) & $\Delta$b($^o$) & d(kpc) & Age & \# of & \# of &
\# of \\  
& & & & Log t & O stars & SNRs & PSRs \\ \hline
SGR OB1 & 4-14 & --2.6 - +1.4 & 1.5 & 6.0-6.6 & 15 & 2 & 4 \\ \\
CYG OB3 & 71-73.8 & +1.2 - +3.4 & & & & & \\
CYG OB1 & 73-79 & --0.6 - +2.84 & 1.2-2 & 6.0-7.2 & 42 & 1 & 0 \\
CYG OB9 & 77-79 & +0.8 - +2.2 & & & & & \\ \\
CEP OB1 & 98-109.3 & --3.1 - +1.3 & & & & & \\
CAS OB2 & 109-114.1 & --1.7 - +3.1 & & & & & \\
CAS OB5 & 114.9-118 & --2.4 - +1.3 & & & & & \\
CAS OB4 & 118-122 & --2.7 - +2.3 & & & & & \\
CAS OB7 & 121.5-125.2 & --0.9 - +2.6 & 1.6-4.8 & 6.0-7.6 & $\sim$80 & 5 & 
5 \\
CAS OB8 & 128-131 & --4 - +2 & & & & & \\
PER OB1 & 132-138 & --5 -\- --2 & & & & & \\
CAS OB6 & 132-142 & --3.7 - +2.3 & & & & & \\ \\
10 OCs & 231-242 & --5.5 - +2 & 1.6-4 & 6.0-7.4 & 10 & 0 & 2 \\  
PUP OB1 & 242-246 & --1 - +2 & & & & & \\ \\
VELA OB2 & 262-268 & --2.7 - +1.4 & & & & & \\
VELA OB4 & 262-268 & --2.7 - +1.4 & 0.6-2 & 6.0-8.0 & 9 & 1 & 1 \\
VELA OB1 & 262-268 & --2.7 - +1.4 & & & & & \\ \\
CAR OB1 & 284.2-286.3 & --2.2 - +0.9 & & & & & \\
9 OCs & 286-289.6 & --1 -\- --0.2 & 2-3.1 & 6.0-7.7 & 84 & 0 & 2 \\
CAR OB2 & 289.2-291.2 & --0.6 - +1.6 & & & & & \\
CRU OB1 & 293.5-295.9 & --2.4 - +0.1 & & & & & \\ \\
CEN OB1 & 301-306.3 & --1.6 - +3.2 & 2.2-2.5 & 6.3-7.5 & 11 & 0 & 1 \\ \\
ARA OB1a & 335.3-336.8 & --1.6 -\- --1.2 & 1.1-2.8 & 6.0-7.4 & 44 & 0 & 2 
\\
13 OCs & 338-346 & --3 - +1.6 & & & & & \\ \\
7 OCs & 351-360 & --2.4 - +1.4 & 1.3-1.7 & 6.8-8.0 & 11 & 0 & 8 \\ \hline

\end{tabular}
\end{flushleft}

\clearpage
\begin{figure}[t]
\vspace{3cm}
\includegraphics{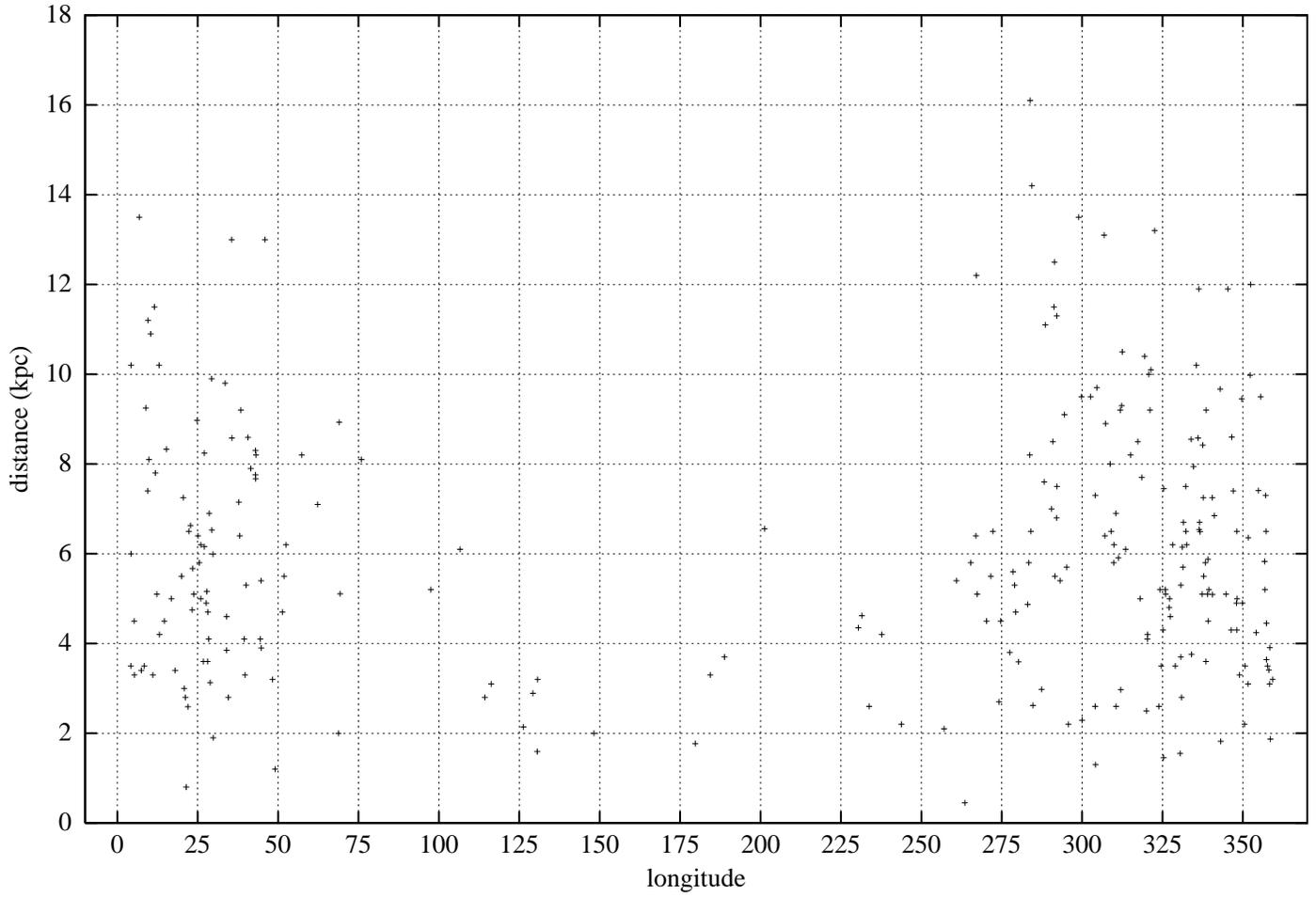}
\caption{Distance versus longitude diagram for 259 PSRs with 
$\mid$b$\mid$$<$5$^o$ and log $\tau$$\le$6.}
\end{figure}

%\clearpage
%\begin{figure}[t]
%\vspace{3cm}
%\special{psfile=background_fig2.ps hoffset=-70 voffset=-150 
%hscale=90 vscale=90 angle=0}
%\caption*
%\end{figure}

%\clearpage
%\begin{figure}[t]
%\vspace{3cm}
%\special{psfile=donuz.ps hoffset=-100 voffset=420 
%hscale=75 vscale=75 angle=270}
%\caption{Dispersion measure vs. longitude diagram for all the 259 PSRs 
%with $\mid$b$\mid$$<$5 and log $\tau$$\le$6}
%\end{figure}

\clearpage
\begin{figure}[t]
\vspace{3cm}
\includegraphics{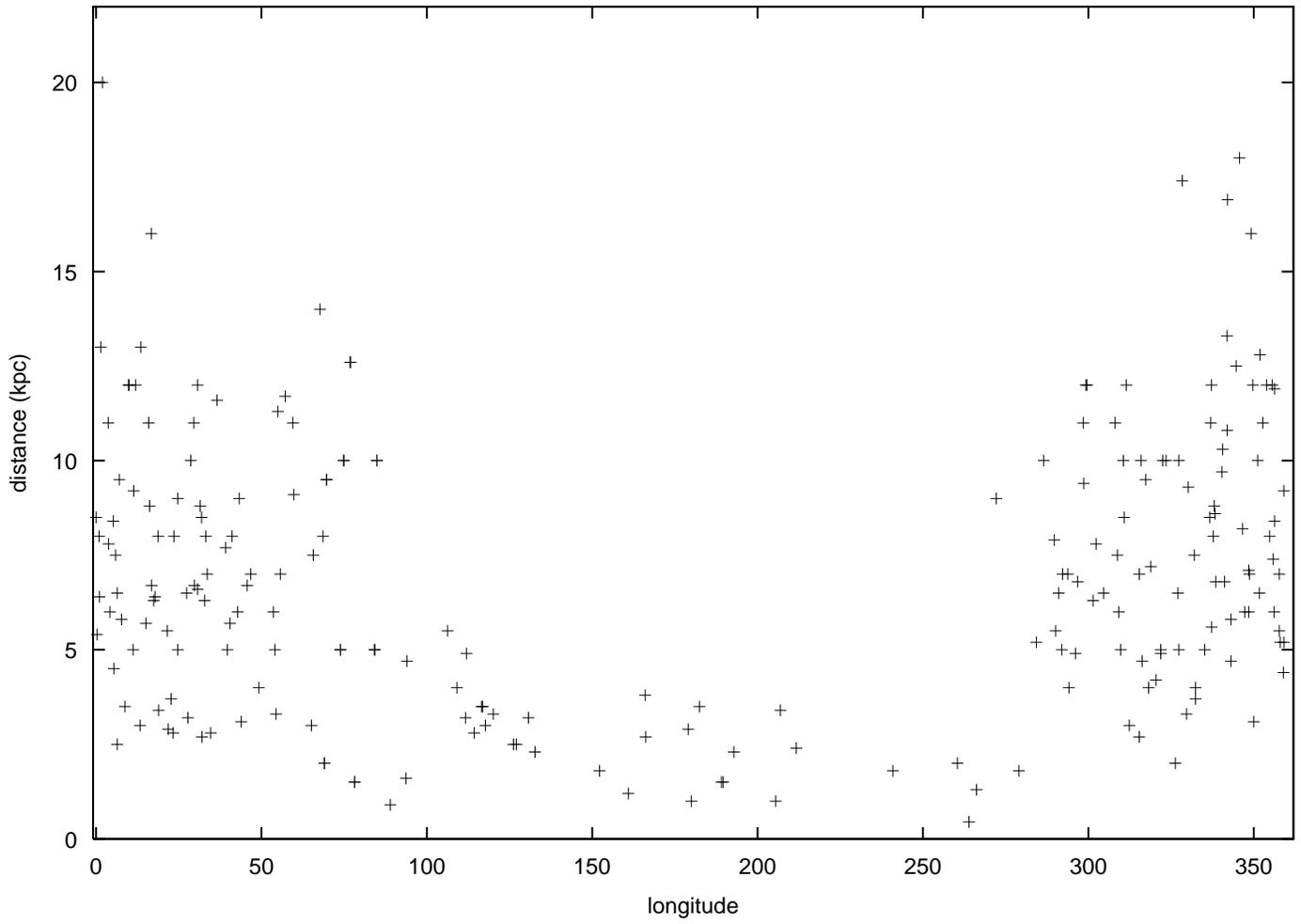}
\caption{Distance vs. longitude diagram for all the 208 SNRs with 
$\mid$b$\mid$$<$5$^o$.} 
\end{figure}

\clearpage
\begin{figure}[t]
\vspace{3cm}
\includegraphics{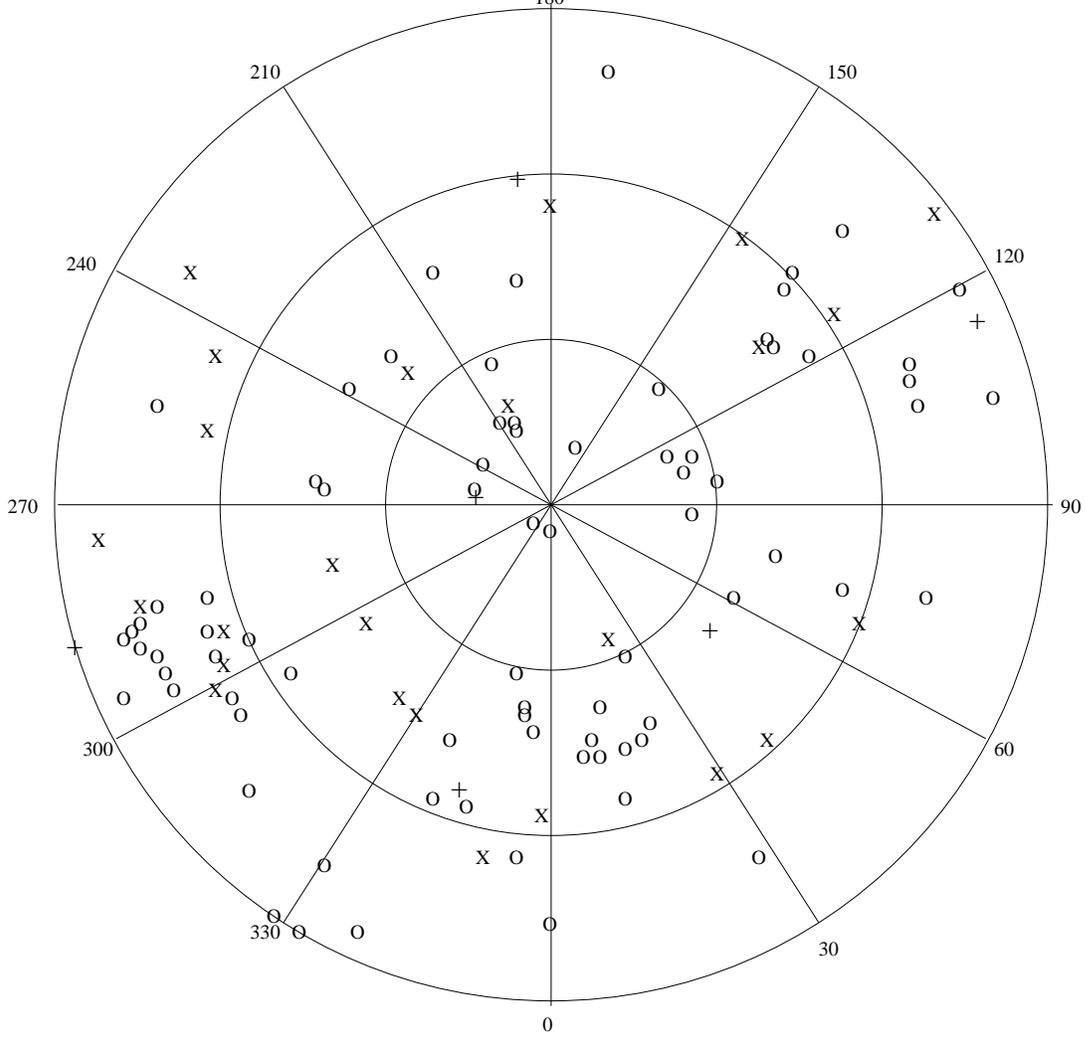}
\caption{Distributions of young PSRs and OB-associations in the Galaxy up 
to 2.5 kpc in the central directions and up to 3 kpc in the anticenter 
directions. PSRs with 10$^5$ $<$ $\tau$ $\le$ 10$^6$ yr are displayed
with sign (X) and PSRs with $\tau$ $\le$ 10$^5$ yr are shown with sign 
(+). Small circles represent OB associations. Three large circles have 
radii 1 kpc, 2 kpc and 3 kpc, respectively, and the Sun is located at the 
center.} 
\end{figure}

\clearpage 
\begin{figure}[t] 
\vspace{3cm}
\includegraphics{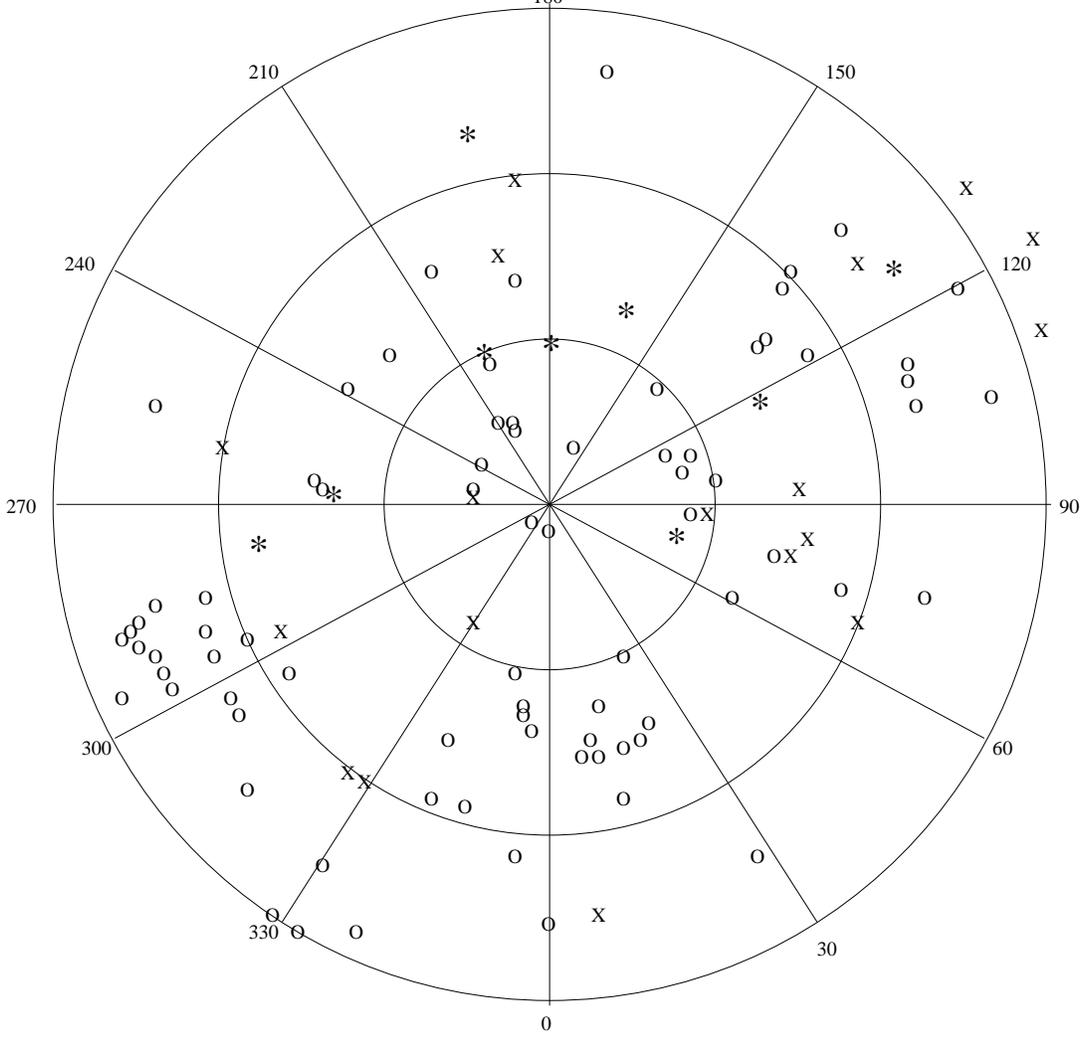} \caption{Distributions of SNRs and OB-associations in
the Galaxy up to 2.5 kpc in the central directions and up to 3 kpc in the
anticenter directions. SNRs with $\Sigma$ $\ge$ 10$^{-21}$
Wm$^{-2}$Hz$^{-1}$sr$^{-1}$ are displayed with sign (X) and SNRs with
3$\times$10$^{-22}$ $\le$ $\Sigma$ $<$ 10$^{-21}$ 
Wm$^{-2}$Hz$^{-1}$sr$^{-1}$ are shown with sign (*). Small circles 
represent OB associations. Three large circles have radii 1 kpc, 2 kpc and 
3 kpc, respectively, and the Sun is located at the center.} 
\end{figure}

\end{document}